\documentstyle[sprocl]{article}

\input{psfig}
\def\veps{\varepsilon}

\def\be{\begin{equation}}
\def\ee{\end{equation}}
\def\bea{\begin{eqnarray}}
\def\eea{\end{eqnarray}}

\begin{document}

\title{CORRELATION--ENHANCED FRIEDEL OSCILLATIONS IN AMORPHOUS
       AND QUASICRYSTALLINE METALS}

\author{JOHANN KROHA}

\address{Institut f\"ur Theorie der Kondensierten Materie,
Universit\"at Karlsruhe, Postfach 6980, 76128 Karlsruhe, Germany}


\maketitle\abstracts{
The exponentially strong damping of the conventional Friedel oscillations
at elevated temperature $T$ as well as due to disorder poses a severe
problem to the Hume--Rothery (HR) stabilization mechanism of amorphous
and quasicrystalline alloys. We show that quantum correlations induced
by electron--electron interactions in the presence of random impurity
scattering can play an important role in stabilizing these systems: When 
there is strong backscattering off local ion clusters, the static electron
density response $\chi (0,q)$ acquires a powerlaw divergence at $q=2k_F$
even at $T>0$. This Fermi surface singularity leads to an enhancement
as well as to a systematical phase shift of the Friedel oscillations, 
consistent with experiments. In addition, the spatial decay exponent is
reduced, strongly supporting the validity of a HR--like mechanism at $T>0$. 
This effect may be accounted for in pseudopotential calculations through 
the local field factor.
}
Many quenched noble--polyvalent metal alloys undergo a crystalline to 
amorphous transformation (CAT) as a function of the noble metal content.
\hbox{Experiments \cite{haeuss.92}} strongly support the 
\hbox{conjecture \cite{beck.79,hafner.79}} that  
the amorphous phase is stabilized by the Hume--Rothery (HR) mechanism, 
i.e.~by forming a structure induced 
pseudogap in the electron density of states (DOS) at the Fermi 
level $\veps_F$. In a real space picture, this is equivalent to
the ions being bound in the minima of the
potential generated by the Friedel oscillations (FOs) of the
conduction electron density
$\rho (r)$ around an arbitrary central ion. The experimental evidence
includes the observation of a pronounced pseudogap, 
a maximum of the electrical resistivity at the CAT and, most
remarkably, the coincidence of the atomic
spacing $a$ with the Friedel wave length
$\lambda _F=\pi / k_F$ near the CAT.  
Very similar behavior is found in icosahedral ({\it i})
\hbox{quasicrystals \cite{davydov.96,stadnik.96,poon.92}}, 
suggesting that the HR mechanism may be active in these systems as well. 
Detailed {\it ab initio} and pseudopotential 
\hbox{calculations \cite{hafner.90,ashcroft.87,fuji.91,hafner.92}} 
have lead to a broad understanding of the features described above. 
However, the HR explanation of the
stability has not unambiguously been accepted due to several unresolved puzzles:
(1) At finite $T$ and also in the presence of disorder the impurity--averaged, 
conventional FOs are exponentially \hbox{damped \cite{degennes.62}}  
due to the spread of the Fermi momentum over a width given by the 
temperature and the inverse elastic mean free path, respectively. Hence,
the structural stability of the strongly disordered amorphous alloys
at elevated temperatures is difficult to
explain by the {\it conventional} FOs alone.
(2) Moreover, the amorphous state is thermally most stable near the
\hbox{CAT \cite{haeuss.92}}.
\begin{figure}
\vspace*{-0.8cm}
\hspace*{2.7cm}{\psfig{figure=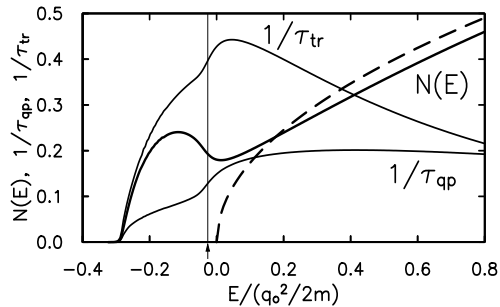,height=4.9cm}}
\caption{Quasiparticle and transport relaxation rates, 
$\tau ^{-1}_{qp}$, $\tau ^{-1}_{tr}$ (units of $q_o^2/2m$), 
and the DOS $N(E)$ (arb. units) as a function of
energy calculated following Ref.~[15] for an impurity scattering T--matrix 
peaked at a momentum transfer $q=q_o$ (see text).  A nearly free electron
model with a bare band
$\veps _p=p^2/2m$ is assumed. ~$- - -$:  clean DOS. Vertical line: position of the Fermi
level
$\veps _F$ for $2k_F=q_o$.
\label{fig:dos}}
\end{figure}
(3) Electron diffraction \hbox{experiments \cite{haeuss.85,haeuss.92}} 
show that in the
amorphous state the ionic positions are systematically shifted compared to the
minima of the Friedel potential of a free electron sea.  
Points (1)--(3) raise the question of a systematical, composition
dependent enhancement and phase shift $\varphi$ of the FOs,
with systematically $\varphi \rightarrow \pi /2$ as the CAT
is approached.
 
The present work is aimed at resolving these problems.
In amorphous alloys the electronic motion is diffusive instead of ballistic.
Diffusion is a dissipative process involving time
reversal symmetry breaking on a macroscopic scale. In a quantum system, like the
electron sea, it arises because the coherence of the wave function is lost on the
scale of the inelastic scattering time, so that the electron motion is
averaged over all possible spatial configurations of the ion system. 
A similar averaging, induced by inelastic
processes, also occurs in {\it real} quasicrystals, so that in these systems the
electrons should be diffusive as \hbox{well \cite{piechon.96}}.  
As will be shown below, the interplay between diffusion and Coulomb interaction can
lead to a strong enhancement and simultaneously to a phase shift of the FOs.
Since diffusion as a dissipative effect is hard to include directly in an {\it ab initio}
calculation, we choose a diagrammatic technique to calculate the dielectric
function $\veps (q)$.

Amorphous \hbox{alloys \cite{haeuss.92}} and {\it i}
\hbox{quasicrystals \cite{poon.92,fuji.91}} are characterized by strong electronic
backscattering off  local concentrical ion clusters,  because the Fermi surface 
nearly coincides  with the spherical Jones or approximant Brillouin zone, respectively; 
i.e.~$2k_F \simeq q_o$, where $q_o$ is the diameter of the Jones zone. This not only
leads to  the  pseudogap in the \hbox{DOS \cite{hafner.90,ashcroft.87,fuji.91}}  but at
the same time \hbox{implies \cite{mahan.90}} an enhancement of the transport
relaxation rate $1/\tau _{tr}$ over the quasiparticle decay rate 
$1/\tau _{qp}$, evidenced by the anomalously large electrical resistivity. The
enhancement is shown in Fig.~\ref{fig:dos}, where both decay rates have been calculated
following Ref.~[15] for a random system  with strong \hbox{backscattering
\cite{kroha.95}}. The pseudogap is also reproduced. Thus, in
amorphous and also in quasicrystalline metals diffusive relaxation generically
dominates over single--particle decay processes, 
so that we may assume  $1/\tau _{tr} \gg 1/\tau _{qp}$. 

\begin{figure}
\centerline{\psfig{figure=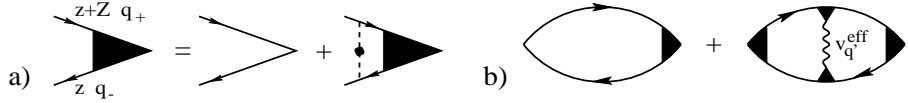,width=\linewidth}}
\caption{a) Diffusion vertex $\Gamma$. b) Polarisation $\Pi (0,q)$
including leading order  quantum correction induced by disorder and interactions.
Dashed lines denote electron--ion scattering, the wavy line with solid
triangles the effective Coulomb interaction.
\label{fig:polaris}}
\end{figure}

We now turn to the calculation of the electronic
density response. 
In a diffusive electron system screening is inhibited, so that the
effective Coulomb interaction $v_q^{eff} (z,Z)$ between electrons with  complex
frequencies $z$ and $z+Z$ acquires a long--range,   retarded \hbox{part
\cite{altsh.79,lee.85}},
\begin{eqnarray}
v_q^{eff}(z,Z)={v_q \over \epsilon ^{RPA}(Z,q)} \Gamma ^2(z,Z,q),\qquad
v_q={4\pi e^2 \over q^2},                 
\label{veff}
\end{eqnarray}
where $\epsilon ^{RPA}(Z,q)\! =\! 1\! +\! 2\pi i\ \sigma\ /
(Z\, \mbox{sgn}Z''+iq^2D)$ is the disordered 
RPA dynamical dielectric function and the diffusion vertex, defined in
Fig.~\ref{fig:polaris} a), is
\begin{eqnarray}
\Gamma (z,Z,q) = \left\{  \begin{array}{ll}
       {i/\tau _{tr} \ \mbox{sgn}Z'' \over  
       Z+iq^2D\ \mbox{sgn}Z'' }\quad & \mbox{if}\ z''(z+Z)''<0  \\
       1 & \mbox{otherwise.}
                          \end{array}
                 \right.                   
\label{vertex}
\end{eqnarray}
$D=1/3\  v_F^2\tau _{tr}$, $\sigma = ne^2\tau _{tr}/m$ 
and $''$ denote the diffusion constant, the Drude
conductivity and the imaginary part, respectively. 
The long--range nature of $v_q^{eff}$ is induced by the 
diffusion pole of $\Gamma$,
which is a consequence of particle number conservation.  Note that it is just
this long--range interaction which causes the $\sqrt{|E-\veps _F|}$ behavior of 
the DOS in disordered systems \cite{altsh.79}, and that the latter has also  
been observed in {\it i} \hbox{quasicrystals \cite{davydov.96}}. 
In order to calculate its effect on the FOs, 
one must consider contributions to the polarisation $\Pi (0,q)$ where 
$\Gamma (z,Z,q')$ enters at $Z,q' \simeq 0$, although $\Pi (0,q)$ is 
evaluated at large external momenta $q\simeq 2k_F$.  
\begin{figure}
\vspace*{-1.2cm}
\centerline{\hspace*{1.2cm}\psfig{figure=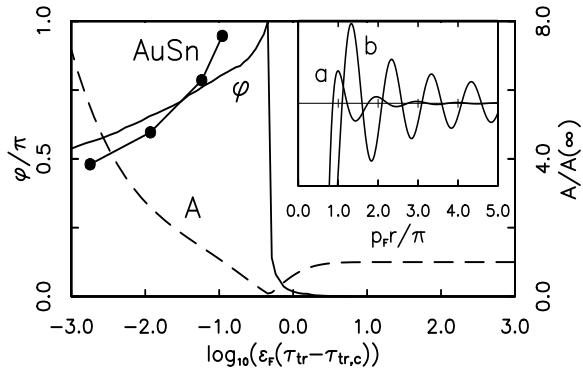,height=5.8cm,width=8.9cm}}
\caption{Phase shift $\varphi$ and amplitude $A$ of the first maximum
of the charge density distribution $\rho (r)$ as a function of the transport time
$\tau_{tr}$ ($T=0.001\veps _F$). 
The inset shows $\rho(r)$ (a) in the clean and (b) in the
strongly disordered ($\veps_F\tau_{tr}=1.23$) case. Data points
represent the phase shift measured [1] in amorphous AuSn, where $\tau_{tr}$
is determined from conductivity measurements up to a constant prefactor.  
\label{fig:osc}}
\end{figure}
The leading singular contribution  
arises from the quantum correction shown in the second diagram
of Fig.~\ref{fig:polaris} b). For amorphous metals 
($1/\tau_{tr}\gg 1/\tau_{qp}$, see above) 
it is evaluated \hbox{as \cite{kroha.95}},
\begin{eqnarray}
\Pi ^{(1)}(0,q) = -{0.280\sqrt{3/2} \over (\varepsilon _F\tau_{tr})^{7/2}}~
{2mk_F\over (2\pi)^2}
\int _{-\veps_F}^{\veps_F} \mbox{d}\nu 
{1/(4T) \over \mbox{cosh}^2{\nu\over 2T}} 
{\mbox{sgn}(x-1)\over \sqrt{|x-1|}},
\label{Pi3}
\end{eqnarray}
where $x=x(\nu )=(q/ 2k_F) / \sqrt{1+\nu / \veps _F}$.
The first term of Fig.~\ref{fig:polaris} b), $\Pi ^{(0)}(0,q)$,
corresponds to the Lindhard function, where 
$\Gamma $ contributes only a nonsingular factor of ${\cal O}(1)$. 
It is seen from Eq.~(\ref{Pi3}) that for $T=0$, 
$\Pi ^{(1)}(0,q)$ exhibits a powerlaw divergence 
$\propto -\mbox{sgn}(q-2k_F)/|q-2k_F|^{1/2}$ at \hbox{$q=2k_F$ \cite{altsh.95}}.
Although at finite $T$ or $1/\tau _{qp}$ the divergence of $\Pi ^{(1)}(0,q)$ 
is reduced to a peak, the inverse dielectric function $1/\veps (q) =1/(1-v_q \Pi (0,q))$
still has a $q=2k_F$ divergence at a critical transport rate $1/\tau _{tr,c}(T)$
{\it even for non--zero single--particle relaxation rate $1/\tau _{qp} <
1/\tau _{tr}$ and at finite $T$}. 
In amorphous systems  the parameter $\tau _{tr}$ may
be varied experimentally by changing the  composition of the alloy. Fourier
transforming $1/\veps (q)$  shows that the peak structure leads to density oscillations
$\rho ^{(1)}(r) \propto \mbox{sin}(2k_Fr)/r^{2}$ superimposed on 
the conventional oscillations $\rho ^{(0)}(r)\propto
\mbox{cos}(2k_Fr)/r^3$, i.e.~to an effective phase
shift $\varphi $, which approaches $\pi /2$ as the weight of the
$2k_F$ peak and, therefore, the amplitude of the oscillations diverges at
$1/\tau _{tr}=1/\tau_{tr,c}$ \hbox{(Fig.~\ref{fig:osc}) \cite{kroha.95}}. 
Near this point the Friedel potential dominates other energy scales present
in the system and can bind the ions in its minima. Note that, in contrast to the
conventional FOs, the divergence of $1 / \veps (q)$ is robust against 
damping due to finite $T$ or disorder. Thus,
identifying the point where the divergence occurs with the CAT explains both the
stability of the amorphous state and the measured, systematical phase
shift $\varphi$ from one single quantum \hbox{effect \cite{kroha.95}}. 
In particular, the fact that the thermal stability of the amorphous
phase reaches its
maximum  near the CAT follows as a natural consequence. 
The structural \hbox{similarities \cite{poon.92}} between amorphous alloys and 
icosahedral quasicrystals and the possible 
presence of a long--range effective interaction
in the quasicrystalline \hbox{systems \cite{davydov.96,altsh.79}}
suggests that the quantum effect discussed above may be important in the latter
systems as well. 
It may be included in pseudopotential calculations through the
local field factor.
\section*{Acknowledgments}
The author has benefited from numerous discussions with 
A.~G.~Aronov, N.~W. Ashcroft, P.~H\"aussler,
A.~Huck, T.~Kopp, Ch.~Lauinger, and P.~W\"olfle, which are gratefully 
acknowledged.
This work is supported by DFG through "SP Quasikristalle".  
Part of the work was performed at LASSP, Cornell University,
supported by a Feodor Lynen Fellowship of the A.~v.~Humboldt Foundation.

\end{document}